\documentclass[12pt]{article}
\usepackage{amssymb}
\input epsf
\newcommand{\be}{\begin{equation}}
\newcommand{\ee}{\end{equation}}
\newcommand{\bea}{\begin{eqnarray}}
\newcommand{\eea}{\end{eqnarray}}
\newcommand{\lb}{\label}
\begin{document}
\begin{titlepage}
\begin{center}
{\large\bf  PRIMORDIAL BLACK HOLES IN AN ACCELERATING UNIVERSE}
\vskip 1cm
{\bf David Polarski}
\vskip 0.4cm
 Laboratoire de Physique Math\'ematique et Th\'eorique,
 UMR 5825 CNRS,\\
 Universit\'e de Montpellier II, 34095 Montpellier, France.\\
\vskip 0.3cm
 Laboratoire de Math\'ematiques et Physique Th\'eorique,
 UMR 6083 CNRS,\\
  Universit\'e de Tours, Parc de Grandmont, 37200 Tours, France.\\
\end{center}
\date{\today}
\vskip 2cm
\begin{center}
{\bf Abstract}
\end{center}

\begin{quote}
General expressions are given for the generation of Primordial Black Holes (PBH) in a universe 
with a presently accelerated expansion due to a(n effective) cosmological constant. 
We give expressions both for a powerlaw scalefree primordial spectrum and for spectra 
which are not of that type.
Specializing to the case of a pure cosmological constant $\Lambda$ and 
assuming flatness, we show that a comological constant with $\Omega_{\Lambda,0}=0.7$ will 
decrease the mass variance at the PBH formation time by about $15\%$ 
compared with a critical density universe. 
\end{quote}

\end{titlepage}
 
\section{Introduction}
The generation of a spectrum of primordial fluctuations in the very Early Universe is the 
crucial ingredient of all inflationary scenarios. These fluctuations can explain the 
generation of all (classical) inhomogeneities that can be seen in our universe, from the Cosmic 
Microwave Background (CMB) anisotropies to the Large Scale Structures (LSS) in the form of 
galaxies and clusters of galaxies. The inflationary paradigm therefore reconciles Big Bang 
cosmology with the appearance of an inhomogeneous universe \cite{L90}. 
In addition, each inflationary 
scenario makes accurate predictions allowing for the observations of ever increasing variety 
and quality to discriminate between the various model candidates. 
One such prediction is the possible formation of Primordial Black Holes (PBH). Indeed, it 
was realized already some time ago that a spectrum of primordial fluctuations would lead to 
the production of PBH \cite{CH74}. 
For this generation mechanism to be efficient, one typically needs a ``blue'' spectrum \cite{CL93}. 
In this way, one can hope that the density contrast averaged over the Hubble radius is sufficiently 
large that the resulting PBH production is not unsignificant and can be used as a powerful 
constraint on the spectrum of inflationary primordial fluctuations and the 
underlying high-energy physics model \cite{GL97,BT01}. 
The production of PBH takes place on scales much smaller than those probed by the CMB anisotropy 
and LSS formation. In this sense, it is analogous if less spectacular, to the generation of 
a primordial gravitational wave background in inflationary models. Of course in the latter case, 
its discovery would be a remarkable prediction of inflation while the existence of PBH is a 
confirmation of the existence of the primordial fluctuations spectrum itself,
irrespective of the way it was generated.
In a recent paper \cite{BKP01}, it was shown that the mass variance 
$\sigma_H(t_k)$ at the PBH formation time $t_k$, 
was significantly overestimated 
due essentially to an incorrect relationship between $\sigma_H(t_k)$ and $\delta_H(t_k)$ 
or equivalently $k^{\frac{3}{2}}~\Phi(k,t_k)$ which quantifies the power at Hubble radius scale 
(these quantities will be carefully introduced below). 
This must be corrected if one is to make accurate predictions. 
In \cite{BKP01}, only a critical density universe was considered and expressions were derived 
for scalefree powerlaw spectra. 
Recent Supernovae observations strongly suggest that we live in a 
presently accelerating universe with $\Omega_{m,0}\approx 0.3,~\Omega_{\Lambda,0}\approx 0.7$, the 
inclusion of a(n effective) cosmological constant seems further 
to make all observations converge into a consistent picture.
Hence, it is important to derive general expressions in our formalism valid for a flat universe 
with an effective cosmological constant like in quintessence models. 
In the particular important case of a pure cosmological constant $\Lambda$, also considered 
in \cite{BK01}, the dependence on $\Lambda$ can be quantified accurately and 
the {\it relative} decrease we find is in agreement with part of the analysis done in \cite{BK01}.
Finally we also generalize our results to primordial perturbations spectra which are not scalefree, 
an interesting possibility to consider in view of the wide range of scales probed by PBH formation.  
We first review the formalism describing PBH formation.    

\section{PBH formation}
We assume for simplicity that a PBH is formed when the density contrast
averaged over a volume of the (linear) size of the Hubble radius
satisfies $\delta_{min} \leq \delta \leq \delta_{max}$,
and further that the PBH mass, $M_{PBH}$, is of the order of the
``horizon mass'' $M_H$, the mass contained inside the Hubble volume. 
Relyimg on semianalytic considerations it is common to take 
$\delta_{min}=\frac{1}{3},~\delta_{max}=1$ but recent numerical simulations 
suggest rather $\delta_{min}\approx 0.7$ \cite{NJ98} and show that $M_{PBH}$ 
can span a certain range, around $M_H$ though, at a given formation time.
More accurately, when some scale defined by the wavenumber $k$ reenters the 
Hubble radius after inflation at some time $t_k$ with $k=(aH)|_{t_k}$, 
it can lead to the production of PBH with $M_{PBH}\approx M_H(t_k)$.
Obviously, there is a one-to-one correspondence between $\frac{a(t_k)}{k}, M_H(t_k)$, 
and $k$.
 
For Gaussian primordial fluctuations, the probability density $p_R(\delta)$, 
where $\delta$ is the density contrast averaged over
a sphere of radius $R$, is given by
\be
p_R(\delta) = \frac{1}{\sqrt{2\pi}~\sigma (R)}~ e^{-\frac{\delta^2}{2
\sigma^2(R)}}\ .
\ee
Here, the dispersion (mass variance) $\sigma^2(R)\equiv \Bigl \langle \Bigl (
\frac{\delta M}{M}
\Bigr )_R^2 \Bigr \rangle$ is computed using a top-hat window function,
\be
\sigma^2(R) = \frac{1}{2\pi^2}\int_0^{\infty}\textrm{d}k\;
 k^2 ~W^2_{TH}(kR) ~P(k)\lb{sigW}~,
\lb{sigma}
\ee
where $P(k)$ is the power spectrum (we assume isotropy of the ensemble). 
From a point of view of principles, the averages 
are quantum averages; however, an effective quantum-to-classical transition is 
achieved during inflation \cite{PS96}. For PBHs produced by 
inflationary perturbations, this quantum-to-classical transition 
is guaranteed for all masses of interest to us (see \cite{P01}).

The expression $W_{TH}(kR)$ stands for the Fourier transform of 
the top-hat window function divided by the probed volume 
$V_W=\frac{4}{3}\pi R^3$,
\be
W_{TH}(kR)=\frac{3}{(kR)^3}\bigl (\sin kR-kR\cos kR\bigr )\ .
\end{equation}

Hence the probability $\beta(M_H)$ that a region of comoving size 
$R=\frac{H^{-1}(t_k)}{a(t_k)}$ has an averaged density contrast at horizon crossing $t_k$ 
in the range $\delta_{min}\leq\delta\leq\delta_{max}$, is given by
\be
\label{beta}
  \beta(M_H)=\frac{1}{\sqrt{2\pi}\,\sigma_H(t_k)}\,
           \int_{\delta_{min}}^{\delta_{max}}\,
           e^{-\frac{\delta^2}{2 \sigma_H^2(t_k)}}\,\textrm{d}\delta
          \approx\frac{\sigma_H(t_k)}{\sqrt{2\pi}\,\delta_{min}}
           e^{-\frac{\delta_{min}^2}{2 \sigma_H^2(t_k)}}\ ,
\ee
where $\sigma_H^2(t_k)\equiv \sigma^2(R)\big|_{t_k}$, and the last approximation
is valid for $\delta_{min}\gg\sigma_H(t_k)$, and
$(\delta_{max}-\delta_{min})\gg\sigma_H(t_k)$.

Important conclusions can be drawn from (\ref{beta}). 
Let us consider first the value of $\beta(M_H)$ 
today. Today we have $\sigma_H^2(t_0)\simeq 10^{-8}$, so clearly the 
probability of forming a black hole 
today is extraordinarily small. This probability
 can increase in the primordial universe if the 
power is increased when we go backwards in time, but the probability  
will remain very small, $\beta(M_H)\ll 1$, at all times due to the magnitude 
of ${\delta^2_{min}}/{\sigma^2_H(t_k)}\gg 1$.

\section{Mass variance in the presence of $\Lambda$}
           \label{secalpha}
When the universe contains a cosmological constant $\Lambda$, this must be taken 
into account for a correct accurate calculation of the mass variance 
at early times. In this section, we will extend the formulas derived in 
\cite{BKP01}. 
As stressed there, one should distinguish the behaviour 
of the quantity $\sigma^2_H(t_k)$, which is ultimately the quantity of 
interest, from the quantity $k^3 \phi^2(k,t_k)$ or $\delta^2_H(k,t_k)$ with 
\be
\delta^2_H(k,t_k)\equiv \frac{k^3}{2\pi^2}~P(k,t_k) = 
          \frac{2}{9 \pi^2}~k^3~\Phi^2(k,t_k)~,\lb{delk}
\ee
where $t_k$ is the PBH formation time of interest, deep in the radiation 
dominated stage. 
However, when dealing with a flat universe with $\Omega_{m,0}<1$, we have {\it today}
\be
\delta^2_H(k_0,t_0)\equiv \frac{k_0^3}{2\pi^2}~P(k_0,t_0) = 
          \frac{2}{9 \pi^2}~\Omega^{-2}_{m,0}~k_0^3~\Phi^2(k_0,t_0)~,\lb{del0}
\ee
where $f_0$ stands for any quantity $f$ evaluated today (at time $t_0$), 
$\Omega_{m,0}=\frac{\rho_{m,0}}{\rho_{cr,0}}$ is the present energy density of 
dust-like matter relative to the critical density and 
$\Omega_{\Lambda,0}\equiv \frac{\Lambda}{3H^2_0}=1-\Omega_{m,0}$.
We first relate the quantities appearing in (\ref{delk},\ref{del0}) at the formation time $t_k$ 
and at the present time $t_0$ for arbitrary evolution of the universe after radiation 
domination and for a scalefree powerlaw spectrum. 
\vskip 10pt
\par\noindent
{\bf General expressions with powerlaw spectrum}: assuming a scalefree powerlaw primordial 
spectrum of the type $k^3 \Phi^2(k)=A(t)~k^{n-1}$ on super Hubble radius (``superhorizon'') scales, 
we then have \cite{PS92}
\be
k^{3}~\Phi^2(k,t_{k}) =  \left (\frac{2}{3} \right )^2~ 
                                \left (1-\frac{H}{a}\int_0^t a~dt' \right)_{t=t_0}^{-2}
~k_0^{3}~\Phi^2(k_0,t_0) \left ( \frac{k}{k_0} \right )^{n-1}~,\lb{phi}
\ee
and analogously
\be
\delta^2_H(k,t_{k}) =  \left (\frac{2}{3} \right )^2~ 
                                 \left(1-\frac{H}{a}\int_0^t a\,dt' \right)_{t=t_0}^{-2}
~\Omega^2_{m,0}~\delta^2_H(k_0,t_0) \left (\frac{k}{k_0} \right )^{n-1}~.\lb{del}
\ee
The lower limit of integration in (\ref{phi},\ref{del}) can be safely taken to be zero.
In (\ref{phi},\ref{del}), we have used a radiation dominated 
stage followed by some {\it arbitrary} evolution of the scale factor. 
In earlier work, we considered a radiation dominated stage followed by a matter 
dominated stage which constitutes a special case of (\ref{phi},\ref{del}). 
An effective cosmological constant as in quintessence models would also be a particular 
case of (\ref{phi},\ref{del}).
However, in contrast to a pure cosmological constant $\Lambda$, the time evolution of the 
scale factor $a(t)$ is model-dependent and cannot be given in full generality. 
The quantity $k_0^{3}~\Phi^2(k_0,t_0)$, or equivalently $\delta^2_H(k_0,t_0)$, at the present 
Hubble radius scale can be derived using the large angular scale CMB anisotropy data. 
It is that quantity that comes from observations which fixes the 
overall amplitude of the fluctuations spectrum. The COBE data show that it is of the following 
order of magnitude \cite{BW97}
\be
k_0^{3}~\Phi^2(k_0,t_0) = 0.86~\times 10^{-8} ~A^2_0(\{n_i\})~,
\ee
where $A^2_0(\{n_i\})$ parametrizes the amplitude variations and is chosen such that 
\be
A_0^2(n=1,\Omega_{m,0}=0.3) \simeq 1~~~~~~~~~~~A_0^2(n=1,\Omega_{m,0}=1) \simeq 1.94~.\lb{A0}
\ee
The exact amplitude depends on the cosmological parameters $\{n_i\}$, 
referring to the background as well as to the inflationary perturbations, and 
this model dependence is encoded in the quantity of order unity $A_0(\{n_i\})$.
Eq.(\ref{A0}) assumes a powerlaw spectrum with spectral index $n$ at least on large scales.
For fixed $n\neq 1$, while the absolute values in (\ref{A0}) are modified, the ratio between 
them is unaltered \cite{BW97}. Note that for a quintessence model, $A^2_0(\{n_i\})$ is model 
dependent. 

Finally, we must relate all our results to the quantity of interest for the computation 
of the PBH abundance, the mass variance $\sigma_H^2(t_k)$ on the Hubble radius scale at 
horizon crossing time $t_k$. As stressed in \cite{BKP01}, one has (with $k=(aH)|_{t_k}$) 
\be
\sigma^2_H(t_k) 
   \equiv \alpha^2(k)~\delta^2_H(k,t_k)~.\lb{alpha1}
\ee
It is crucial to distinguish both quantities $\sigma_H(t_k)$ and $\delta_H(k,t_k)$. 
As seen from (\ref{alpha1}), the quantity $\sigma_H^2(t_k)$ which depends on the 
averaging procedure through eq.(\ref{sigW}) is correctly related to 
the (non-averaged) quantity $\delta^2_H(k,t_k)$ in a non-trivial way.  
The quantity $\delta_H(k,t_k)$ can be reconstructed at the time $t_k$ from its present value 
$\delta_H(k_0,t_0)$ using (\ref{del}).
But this is {\it not} the case for the quantity 
$\sigma_H(t_k)$ because the deformation of the power spectrum is different at the 
time $t_k$ and today. In other words, 
\be
T(k',t_0)\neq T(k',t_k)~,\lb{T0}
\ee 
where the transfer function $T(k,t)$ is defined through
\be\lb{T}
P(k,t)= \frac{P(0,t)}{P(0,t_i)}~P(k,t_i)~T^2(k,t)\; , \quad T(k\to 0,t)\to 1~.
\ee
Here, $t_i$ is some initial time when all scales are outside the Hubble radius, 
$k\ll aH$, we can take for example $t_{i}=t_e$, the end of inflation.
For a powerlaw scalefree spectrum we have \cite{BKP01}
\be
\alpha^2(k)= \int_0^{\frac{k_e}{k}} x^{n+2}~
 T^2(kx,t_k)~W^2_{TH}(x)~\textrm{d}x~,\lb{alpha}
\ee
where $k_e$ corresponds to the shortest fluctuations wavelengths with the size of the 
Hubble radius at the end of inflation. 
The transfer function $T(k',t)$ in the integrand of (\ref{alpha}) must be taken 
{\it at the time $t_k$}, not today. 
The accurate value of $\alpha(k)$ requires numerical calculations but estimates 
made in \cite{BKP01} show a significant overestimation of the mass variance $\sigma_H(t_k)$ 
when it is not computed correctly using the right quantity $\alpha(k)$. One then gets that 
$(10/9)^2\alpha^2(k)\ll 25$ (the value usually taken in the literature for a 
critical density universe) for all the mass 
range of PBH produced in the radiation era. If one is willing to use PBH formation as a 
precision tool in cosmology, it is important to check in how far the presence of a cosmological 
constant with $\Omega_{\Lambda,0}=0.7$ brings further modifications.

An important conclusion can be immediately drawn by inspection of the integrand in (\ref{alpha}) 
without accurate knowledge of the transfer function at time $t_k$. Indeed, as we are interested 
in times $t_k\ll t_{eq}$ and in universes where $\Omega_{\Lambda}$ 
domination occurs late, it is clear that at the time $t_k$, neither the long-wave nor the 
short-wave fluctuation modes are affected in any way by the presence of  
(an effective) $\Lambda$. As can be seen from (\ref{T}), this implies that the transfer function 
at time $t_k$ does not depend on $\Lambda$ and the same must apply therefore to the quantity 
$\alpha(k)$. 
We conclude that the influence of a(n effective) cosmological constant 
on the probability $\beta(M_H)$ comes solely from its influence on the quantity 
$\delta^2_H(k,t_k)$, or $k^{3}~\Phi^2(k,t_{k})$. It is this influence that we will quantify in the 
next subsection. We now consider a powerlaw spectrum and specialize to a universe with a 
cosmological constant $\Lambda$.
 
\vskip 10pt
\par\noindent 
{\bf Powerlaw spectrum with ${\bf \Lambda}$}: in order to account for the presence of a 
cosmological constant $\Lambda$, we must replace the evolution of the scale factor $a(t)$ after 
the radiation dominated stage. The scale factor for this stage of the universe evolution is 
very well approximated by \cite{SS00}   
\be
a(t)=a_1 \sinh^{2/3}(\beta t)~,\lb{a}
\ee
where $\frac{2}{3}\beta =\sqrt{\Lambda/3}=H_0 \sqrt{\Omega_{\Lambda,0}}$, 
$a_1=$const.
The evolution (\ref{a}) smoothly interpolates between a pure (flat) dust-like matter dominated 
stage, with $a(t)\propto t^{\frac{2}{3}}$, for $\beta t\ll 1$ which is of course 
the case right after $t_{eq}$, and a $\Lambda$ dominated universe in the asymptotic future.  
In particular for scales for which $z_{eq}>z(t_k)\gg 1$, $a(t_k)\propto t_k^{\frac{2}{3}}$.
It is this evolution (\ref{a}) which must be used in (\ref{phi},\ref{del}). 
It is physically appealing to express the results in terms of the quantity $M_H(t_k)$. 
Then the following result is obtained 
\bea
k^{3}~\Phi^2(k,t_{k}) &=&  \left (\frac{2}{3} \right )^2~ 
                                \left (1-\frac{H}{a}\int_0^t a~dt' \right)_{t=t_0}^{-2}
~k_0^{3}~\Phi^2(k_0,t_0)~\nonumber\\
&\times & \Biggl \lbrack  \frac{k_{eq}}{k_0}  \Biggr  \rbrack^{n-1}~
\Biggl \lbrack  \frac{M_H(t_{eq})}{M_H(t_k)}  \Biggr \rbrack^{\frac{n-1}{2}}\lb{Lam}~.
\eea
The evolution (\ref{a}) must now be substituted in (\ref{Lam}).
Actually it is slightly more accurate to write the quantity $\frac{k}{k_0}$ as 
\be
\frac{k}{k_0}\propto M_H(t_k)^{-\frac{1}{2}}~,~~~~~~~~~~~~~t_k\ll t_{eq}\lb{prop}
\ee
where the proportionality constant is {\it independent} of $\Omega_{m,0}$ 
(and $\Omega_{\Lambda,0}$). 
We will take now the following quantities: 
$\Omega_{r,0}= 9.81 \times 10^{-5}~\frac{g_{eff}}{3.36}~h_{65}^{-2}~(\frac{T_{\gamma,0}}{2.726})^4$ 
is the relative energy density of relativistic matter today, 
$g_{eff}$ is the present effective number of relativistic degrees of freedom, 
$h_{65}\equiv \frac{H_0}{65} {\rm km/s/Mpc}$, $T_{\gamma,0}$ is the present temperature of the CMB.
Using these quantities, eq.(\ref{Lam}) finally becomes for $\Omega_{\Lambda,0}=0.7$
\bea
k^{3}~\Phi^2(k,t_{k}) &=&  1.75 \times 10^{-8}
                 ~\frac{0.219}{I^2(\Omega_{\Lambda,0})}
                    ~A^2_0(\{n_i\})\nonumber\\
&\times & \Biggl \lbrack 9.75 \times 10^{26}~h_{65}^{-1}~\left( \frac{g_{eff}}{3.36}\right )^{\frac{1}{4}}
 \frac{T_{\gamma,0}}{2.726 ^{\circ}K} \Biggr \rbrack^{n-1}
\Biggl \lbrack  \frac{\rm g}{M_H(t_k)}  \Biggr \rbrack ^{\frac{n-1}{2}}\lb{Lam1}~,
\eea
where $M_H(t_k)$ is expressed in grams.
In deriving (\ref{Lam1}) we have used 
\bea
I(\Omega_{\Lambda,0}) &\equiv&  1-\frac{H_0}{a_0}\int_0^{t_0} a(t)~dt \simeq 
1-\frac{2}{3\sqrt{\Omega_{\Lambda,0}}} 
        \int_0^d \Biggl \lbrack \frac{\sinh x}{\sinh d}\Biggr \rbrack ^{\frac{2}{3}} dx\\
d &\equiv& \frac{1}{2} \ln \frac{1+\sqrt{\Omega_{\Lambda,0}}}{1-\sqrt{\Omega_{\Lambda,0}}}~,
\eea
which gives the following numerical result substituted in (\ref{Lam1})
\be
I(0.7) = 0.468~.\lb{int0.7} 
\ee
Hence, $I^{-2}(0.7)=4.57$ gives an increase of $64\%$ compared to 
the value $25/9=2.78$ obtained for a critical density flat universe, $\Omega_{m,0}=1$. 
On the other hand normalization to the CMB fluctuations decreases $A^2_0$ by a little bit more 
than $48\%$, as seen from (\ref{A0}). Therefore both effects combined lead to a decrease 
of about 15\%.  
As for $\alpha(k)$, we have seen above that it is insensitive to the 
presence of a cosmological constant. 
From (\ref{prop}), the correspondence between the (approximate) PBH mass $M_H(t_k)$ and $k$ 
is independent of $\Omega_{m,0}$ for $t_k\ll t_{eq}$.
Therefore we conclude that, in a flat universe, a cosmological 
constant with $\Omega_{\Lambda,0}=0.7$ will decrease the mass variance $\sigma^2_H(t_k)$ as follows
\be
\sigma^2_H(t_k)|_{\Omega_{\Lambda,0}=0.7} \simeq 0.85 ~
                           \sigma^2_H(t_k)|_{\Omega_{m,0}=1}\lb{dec}
\ee
As a result, the significantly lower value estimated in \cite{BKP01} is further reduced by about 
$15\%$, further diminishing the probability for PBH formation. 
This is in good agreement with the analysis performed in \cite{BK01}.
Still, it is instructive that these authors have incorrectly assumed that 
$\sigma_H(t_k)=C~\delta_H(t_k)$ where the constant 
$C$ is the same at all times $t_k$ up to the present time $t_0$. Nevertheless, their 
eq.(4.30) would give essentially the same {\it relative} decrease of $\sigma_H(t_k)$ 
due to the presence of a cosmological constant as in (\ref{dec}). This is because the correct factor 
$\alpha(k)$ does not depend on $\Lambda$.

We finally note that equation (\ref{Lam1}) makes use of the observed amplitude today on the 
present Hubble radius scale and, as far as perturbations are concerned,
combines it with an {\it assumed} (powerlaw) behaviour towards small scales on 
a very broad range of scales. Other behaviours are certainly possible however. 
This is reminiscent of the primordial gravitational wave background generated during inflation 
extending up to frequencies as high as $10^{10}$Hz. There too, one can imagine a behaviour towards 
large frequencies departing from a simple scalefree law (see e.g. \cite{P99} for such a model 
with a jump in the tensorial spectral index $n_T$). 
Hence if the assumption of a scalefree spectrum does not hold, a more general expression will 
be needed. For this reason, we now generalize our results also to these cases.

\vskip 10pt
\par\noindent 
{\bf Spectrum with a characteristic scale}:
A further important generalization concerns the primordial fluctuations spectrum 
itself. Indeed, the equations written in the previous subsection assume a scalefree 
spectrum. However, this needs not be the case especially in view of the large range 
of scales that are probed by PBH formation. 

Let us therefore define in full generality  
\be
k^{3}~\Phi^2(k,t_{k}) \equiv \left (1-\frac{H}{a}\int_0^t a~dt' \right)_{t=t_k}^2~F(k) 
                                         = \frac{4}{9}~F(k)~,\lb{F} 
\ee
where $F(k)$ can be any complicated function of $k$. The expression (\ref{F}) represents the primordial 
spectrum on ``super-Hubble radius'' (superhorizon) scales. 
For example, the spectrum of double inflation considered in \cite{PS92,P94} is of this general type. 
The corresponding generalization of 
(\ref{phi}), or (\ref{del}), leads to a more complicated equation, viz.
\bea
k^{3}~\Phi^2(k,t_{k}) &=&  \left (\frac{2}{3} \right )^2~ 
                                \left (1-\frac{H}{a}\int_0^t a~dt' \right)_{t=t_0}^{-2}
~k_0^{3}~\Phi^2(k_0,t_0)\nonumber\\ 
&\times& 
\frac{F(\alpha_1 M_H^{-\frac{1}{2}}(t_k))}{F(k_0)}~,
\lb{phiF}
\eea
where $\alpha_1$=constant.
In case the function $F(k)$ is actually of the form $F(\frac{k}{k_s})$, where $k_s$ 
defines the characteristic scale -- an example of such a spectrum was found in \cite{S92} 
and considered in \cite{LPS98} -- eq.(\ref{phiF}) can be recast into a slightly simpler form
\bea
k^{3}~\Phi^2(k,t_{k}) &=&  \left (\frac{2}{3} \right )^2~ 
                                \left (1-\frac{H}{a}\int_0^t a~dt' \right)_{t=t_0}^{-2}
~k_0^{3}~\Phi^2(k_0,t_0)\nonumber\\
&\times& \frac{F\left (\sqrt{\frac{M_s}{M_H(t_k)}} \right )}
                              {F(k_0/k_s)}\lb{bsi}
\eea
with $M_s\equiv M_H(t_{k_s})$ and $t_{k_s}<t_{eq}$. 
Specializing to the particular case $\Omega_{\Lambda,0}
=0.7$ just requires the substitution, like in (\ref{Lam1}), of the corresponding numbers into 
(\ref{phiF},\ref{bsi}). 

Finally, we come to the calculation of $\sigma_H(t_k)$ itself. We now have the corresponding 
generalization to primordial spectra of arbitrary shape
\be
\sigma^2_H(t_k) = \frac{8}{81 \pi^2}~
 \int_0^{\frac{k_e}{k}} F(kx)~x^{3}~T^2(kx,t_k)~W^2_{TH}(x)~\textrm{d}x~.
\ee
In particular, the general expression for $\alpha^2(k)$ is given by 
\be
\alpha^2(k)= \int_0^{\frac{k_e}{k}} \frac{F(kx)}{F(k)}~
x^{3}~T^2(kx,t_k)~W^2_{TH}(x)~\textrm{d}x~.\lb{alphaF}
\ee
Again, for the case of interest to us, $t_k\ll t_{eq}$, the transfer function at the time $t_k$ and 
therefore also $\alpha(k)$ are independent of $\Lambda$. We conclude that the same decrease found 
in (\ref{dec}) will apply here too.
Note that (\ref{alphaF}) extends the result (\ref{alpha}) derived in \cite{BKP01} for 
a powerlaw scalefree spectrum which just corresponds to $F(k)\propto k^{n-1}$.
A characteristic scale in the primordial spectrum is an interesting possibility 
with respect to PBH formation in view of the large range of scales involved, much larger than CMB 
anisotropy or LSS formation. 
We have already considered some simple toy models in \cite{BKP01} and interesting results were 
obtained. The detailed numerical study of more sophisticated spectra and their 
possible relevance to observations is under progress \cite{BBKP02}.

%\section*{Acknowledgements}

\end{document}